# Direct phase-locking of a Ti:Sapphire optical frequency comb to a remote optical frequency standard

EUNMI CHAE,[1] KOTA NAKASHIMA,[2] TAKUYA IKEDA,[2] KEI SUGIYAMA,[2,3] AND KOSUKE YOSHIOKA[1,*]

[1]*Photon Science Center, School of Engineering, The University of Tokyo,*
*2-11-16 Yayoi, Bunkyo-ku, Tokyo 118-8656, Japan*
[2]*Department of Applied Physics, School of Engineering, The University of Tokyo,*
*7-3-1 Hongo, Bunkyo-ku, Tokyo 113-8656, Japan*
[3]*Currently with the Institute for Solid State Physics, The University of Tokyo,*
*5-1-5 Kashiwanoha, Kashiwa, Chiba 277-8581, Japan*
*\*yoshioka@fs.t.u-tokyo.ac.jp*

**Abstract:** We report on an ultralow noise optical frequency transfer from a remotely located Sr optical lattice clock laser to a Ti:Sapphire optical frequency comb through telecom-wavelength optical fiber networks. The inherent narrow linewidth of the Ti:Sapphire optical frequency comb eliminates the need for a local reference high-finesse cavity. The relative fractional frequency instability of the optical frequency comb with respect to the remote optical reference was $6.7(1) \times 10^{-18}$ at 1 s and $1.05(3) \times 10^{-19}$ at 1,000 s including a 2.9 km-long fiber network. This ensured the optical frequency comb had the same precision as the optical standard. Our result paves the way for ultrahigh-precision spectroscopy and conversion of the highly precise optical frequency to radio frequencies in a simpler setup.

## 1. Introduction

Optical frequencies have been the most precisely measured quantities since the development of optical frequency combs (OFCs) [1]. With the unprecedented precision, optical precision spectroscopy has been used for examining the theory of quantum electrodynamics [2, 3], space and time variations of fundamental constants [4–6], relativistic geodesy [7, 8] and the existence of dark matter [9, 10] using its ability to detect infinitesimal energy shifts due to these exotic effects. The heart of such high-precision optical spectroscopy is high frequency stabilities and accuracies of ultranarrow light sources at desired wavelengths.

A continuous wave laser (CW laser) stabilized to a high-Q cavity is often employed as a light source for the precision optical spectroscopy [11, 12]. The state-of-the-art CW laser stabilized to a silicon Fabry-Pérot cavity at 124 K has accomplished a sub-10 mHz linewidth and coherence time longer than 10 s [11]. Optical atomic clocks, considered as the next-generation of the time and frequency standards [13–19], have achieved a precision of $3.1 \times 10^{-17}/\sqrt{\tau}$ and $5 \times 10^{-19}$ in 1 h [16] by utilizing these ultranarrow linewidth lasers.

However, preparing the state-of-the-art optical references at each laboratory requires tremendous resources and efforts. Instead, the coherence and the instability of the optical references can be shared by delivering the optical references to remote laboratories. Currently, the best way to distribute optical standards to remote places is through optical fiber networks [20–29]. Fiber-induced phase noise can be actively cancelled out (fiber noise cancelling; FNC), and it has already proven that the stabilities and accuracies of the optical standards [21, 30] are conserved when the light is transmitted through the fiber networks. However, the FNC is never effective for the phase fluctuations whose time scales are faster than the round-trip propagation time ($2T$) through the fiber; the FNC leaves unwanted phase noises appearing above the feedback bandwidth of about $1/4T$. Therefore, a local ultranarrow CW laser is commonly employed and is locked to the coherent portion of the transferred optical standard.

Femtosecond OFCs have played pivotal roles in promoting precision optical spectroscopy due to their capability of measuring the absolute frequency of laser light [31–38]. An important feature of OFCs is the coherence transfer of optical frequencies across hundreds of THz. The conversion of the received optical standard to desired wavelengths can be accomplished by phase-locking an OFC to the received optical reference [39]. Due to the remaining high frequency phase noises from the fiber networks, an appropriate combination of an OFC and a ultranarrow CW laser is usually necessary at the receiver site to filter out the fiber-induced high frequency phase noise and to transfer the coherence of the optical standard to other wavelengths. However, it is possible to directly phase-lock the local OFCs against the remote optical reference without using the local ultranarrow CW laser, provided that the employed fiber length is moderate and that the optical linewidth of the free-running optical frequency comb is sufficiently narrow. This can greatly reduce the complexity accompanied by the use of a high-finesse cavity at the receiver site.

In this work, we demonstrate a coherent transfer of the remote optical reference at 698 nm directly to a Ti:Sapphire OFC through a 2.9 km-long commercial optical fiber network. The Ti:Sapphire OFC has an inherently narrow linewidth of on the order of 10 kHz, which is approximately two orders of magnitude narrower than fiber-based optical frequency combs. This narrow linewidth assures direct transferring of the linewidth of the optical reference without additional high frequency phase noise when the length of the optical fiber networks is of the order of km, which corresponds to an FNC bandwidth of 10s of kHz. We envision the extension of this demonstration to a future setup that enables us to conduct ultrahigh-precision spectroscopy in the visible and ultraviolet regions just by connecting a telecommunication fiber to a Ti:Sapphire OFC.

## 2. Experimental setup and results

### 2.1 Optical fiber link

The overall experimental setup is depicted in Fig. 1. The experiment was conducted at two laboratories at the University of Tokyo: Katori laboratory at Hongo campus (Site1) and Yoshioka laboratory at Asano campus (Site2). As an optical frequency standard, an external cavity diode laser (ECDL) at 698 nm for the Sr optical lattice clock (the clock laser) with a stability of $2\times10^{-15}$ at 1 s was used [40]. The performance of this experiment was evaluated using residual phase noises and fractional frequency instabilities which indicate how coherently the stability of the clock laser at Site1 was transferred to the Ti:Sapphire optical frequency comb at Site2. The clock laser was sent through a phase noise-compensated polarization-maintaining single mode optical fiber (FNC1) to where the first optical repeater (Repeater1) was located. Repeater1, an ECDL at 1,397 nm, was frequency-doubled to produce 698 nm, which is phase-locked to the delivered clock laser using a heterodyne beat. The remaining 1,397 nm light was used for the transmission from Site1 to Site2 through a dark fiber of the optical fiber network at the University of Tokyo (UTNET), which connects all buildings in the university campus using telecom fibers and multi-channel switches. The total length of the optical fibers used in this study was approximately 2.9 km. The corresponding feedback bandwidth for FNC is approximately 17 kHz. The received power of the repeater laser was about 20 % of the transmission power because of insertion losses at more than ten switches and fiber connections within the path. In order to overcome the power loss, a second ECDL at 1,397 nm (Repeater2) was built to copy Repeater1 by phase-locking the heterodyne beat. Part of the 1,397 nm light from Repeater2 was sent back through UTNET and used to cancel the phase noise of the fiber network between Site1 and Site2 (FNC2). The remaining 1,397 nm light was frequency-doubled to 698 nm for the stabilization of the optical frequency comb.

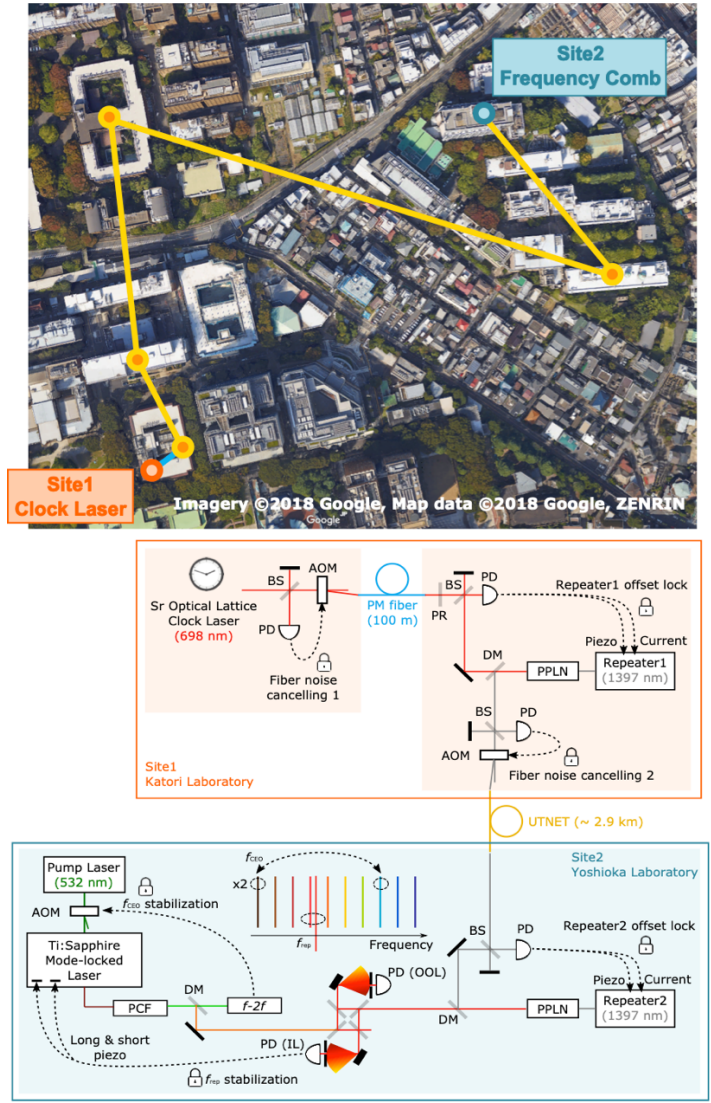

Fig. 1. Overall experimental setup. The optical standard, a Sr optical lattice clock laser, was coherently transferred using a 2.9 km-long optical fiber network. Two phase-locked ECDLs were installed for conversion between the wavelength of the optical standard (698 nm) and a near-telecom wavelength of 1,397 nm for transmission efficiency. A Ti:Sapphire optical frequency comb was used as a frequency synthesizer. $f_{CEO}$ and the heterodyne beat between one tooth of the frequency comb and the optical standard were phase-locked. All RF signals and detectors (counters and spectrum analyzers) were referenced by a GPS-disciplined Rb standard. PM fiber: polarization-maintaining fiber; PD: photodetector; BS: beam splitter; DM: dichroic mirror; PR: partial reflector; PPLN: periodically poled lithium niobate; AOM: acousto-optic modulator; IL: in-loop; OOL: out-of-loop, Image from Google.

The spectrum of the phase-locked heterodyne beat signals between the clock laser and Repeater1 and between Repeater1 and Repeater2 are presented in Figs. 2(a) and 2(b) and Figs. 2(c) and 2(d), respectively. The measured linewidth of 1 Hz in narrower spectral window (Figs. 2(b) and 2(d)) was limited by the instrumental resolution. The narrow peak at the center of the spectrum indicated the phase-coherent component of the repeaters to the clock laser. Due to the relatively broad spectrum of the free-running repeater lasers, the spectra had shoulders at 500 kHz upto 3 MHz away from the center frequency after frequency noises lower than 500 kHz were covered by the feedback loops (Figs. 2(a) and 2(c)). This remaining broad spectrum feature does not affect the stabilization of the OFC since the Ti:Sapphire OFC has an intrinsically narrow linewidth of on the order of 10 kHz.

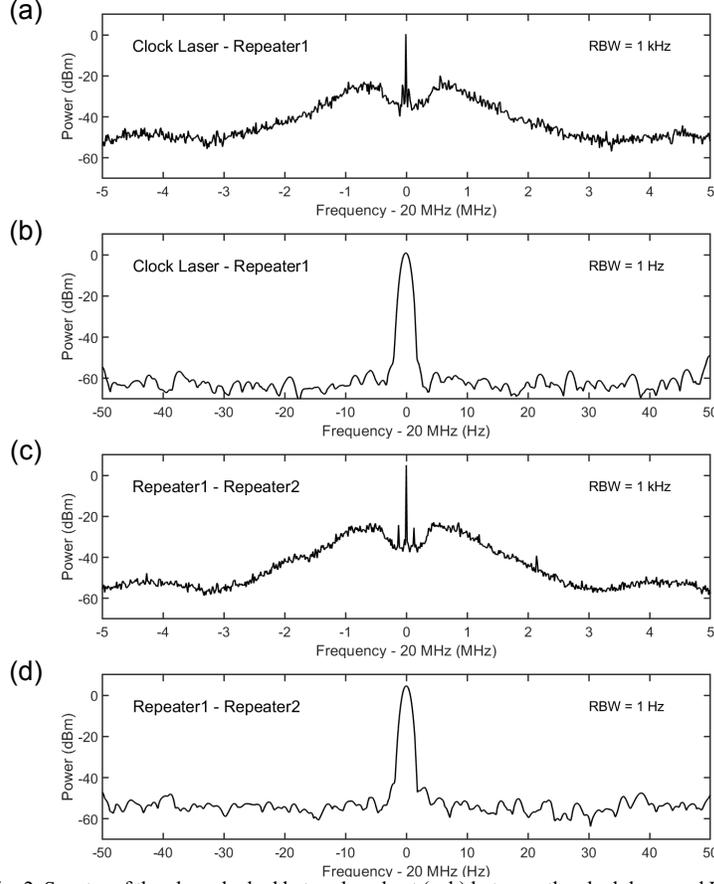

Fig. 2. Spectra of the phase-locked heterodyne beat (a, b) between the clock laser and Repeater1 and (c, d) between Repeater1 and Repeater2. (a, c) Spectra with a 10 MHz span with the resolution bandwidth (RBW) of 1 kHz. (b, d) Spectra with 100 Hz span with 1 Hz RBW. The observed linewidth of the coherent component was limited by the instrumental resolution.

To examine the low frequency part of the beat signal in detail, we measured the power spectral density (PSD) of the residual phase noises at each part of the optical fiber link (Fig. 3) with the resolution of 8.3 mHz. Since the one-way fiber noise cannot be perfectly suppressed when the round-trip fiber noise is cancelled with a phase-locked loop (PLL) [30], we followed and extended the analysis of Ref. 30 to estimate the residual one-way fiber noises after the round-trip PLL. The phase noise PSDs for the FNCs depicted in Fig. 3 are the upper limit of the residual one-way fiber noises estimated from the free-running round-trip phase noise, the residual phase noise of the PLL. The in-loop phase noise of the Repeater2 was also taken into account for FNC2 as well. The detailed methods are described in Appendices 4.2. All phase noise PSDs show $1/f$ tendency at low frequencies below 1 Hz, indicating flicker phase modulation from the PLL. This is due to the fact that the phase noises of PLLs are dominant contributions to the residual one-way fiber noises at low frequencies. The phase accumulations at each part are summarized in Table 1. The total accumulated phase from 8.3 mHz to 195 kHz due to the fiber network was 1.0227 rad.

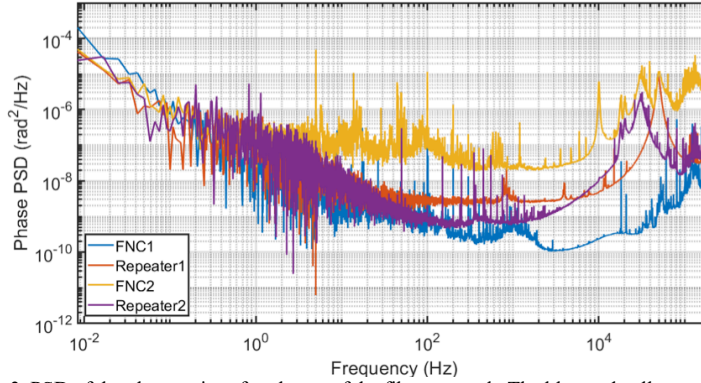

Fig. 3. PSD of the phase noise of each part of the fiber network. The blue and yellow traces show the upper limit of the residual one-way fiber noise of FNC1 and FNC2 respectively after cancelling the round-trip fiber noise. The scarlet(purple) trance is the phase noise of the in-loop heterodyne beat between the optical standard and Repeater1(Repeater1 and Repeater2).

**Table 1: Residual phase accumulation at each part of the fiber network.**

|  | Phase accumulation (8.3mHz ~ 195 kHz) |
|---|---|
| FNC1 | 0.0463 rad |
| Repeater1 | 0.2479 rad |
| Repeater2 | 0.1665 rad |
| FNC2 | 0.9770 rad |
| **Total** | **1.0227 rad** |

## 2.2 Phase-locking of an OFC

A home-made, femtosecond mode-locked Ti:Sapphire oscillator was chosen to construct the OFC that converts the optical frequency standard to the visible and near-infrared wavelengths. The spectrum of the OFC spanned from 700 nm to 900 nm with an output power of 400 mW when pumped by a single-mode green laser at 3.2 W. The oscillator could produce an output power of 950 mW at a pump power of 5 W. The repetition rate ($f_{rep}$) of the laser was 120 MHz. About 30% of the output power was used for coupling to a commercial photonic crystal fiber (PCF) with an efficiency of 60% for generating an octave-spanning spectrum from 500 nm to 1,100 nm. The carrier-envelope offset frequency ($f_{CEO}$) was actively stabilized to 11.01 MHz using the self-referencing technique with an $f$-$2f$ interferometer by adjusting the pump laser power using an acousto-optic modulator (AOM). The inherent narrow linewidth of the $f_{CEO}$ spectrum of the home-built Ti:Sapphire frequency comb contributed to the exceptionally low instability of the frequency comb achieved in this work (see Appendices 4.3).

The remaining degree of freedom of the optical frequency comb, $f_{rep}$, was stabilized directly by referencing the delivered optical standard. Phase-locking one tooth of the optical frequency comb to the second-harmonic generation (SHG) of Repeater2, which is a coherent copy of the optical standard clock laser, resulted in precise stabilization of the frequency comb's $f_{rep}$ when $f_{CEO}$ is fixed. Part of the light after the PCF, with wavelengths longer than 550 nm, was combined with the SHG of Repeater2 using a beam splitter. After a diffraction grating to discard unnecessary wavelengths, a heterodyne beat between one tooth of the optical frequency comb and the clock laser was measured by a fast photodetector (PD). Stabilization of the heterodyne beat to 10.01 MHz was accomplished by applying feedback voltages to piezoelectric transducers (PZTs) inside the Ti:Sapphire oscillator to adjust the cavity length of the oscillator.

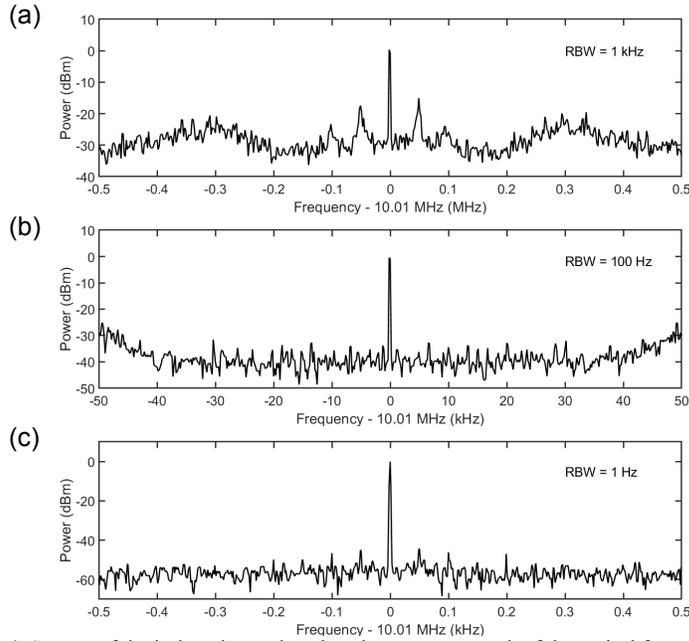

Fig. 4. Spectra of the in-loop heterodyne beat between one tooth of the optical frequency comb at 698 nm and the SHG of Repeater2 when the comb is phase-locked. (a) 1 MHz span with 1 kHz RBW. (b) 100 kHz span with 100 Hz RBW. (c) 1 kHz span with the RBW of 1 Hz. The phase-coherent component was observed with a signal-to-noise ratio of more than 50 dB.

The power spectrum of the phase-locked heterodyne beat between the copy of the clock laser and a tooth of the Ti:Sapphire optical frequency comb was measured by a spectrum analyzer (Fig. 4). The resolution-limited sharp feature in the center confirmed that the tooth of the OFC was phase-coherent to the copy of the optical standard clock laser. In addition, the PSD of the phase noise in both in-loop and out-of-loop heterodyne beat signals between the frequency comb and Repeater2 were obtained independently with the resolution of 8.3 mHz and 160 kHz cutoff bandwidth (Fig. 5). The cutoff frequency was chosen as 160 kHz since the noise components with frequencies higher than that are solely from the pedestal phase noise of Repeater2. To minimize the sensitivity to the in-loop and out-of-loop optical path fluctuation, the uncommon optical path was constructed as short as possible. The in-loop and out-of-loop measurements agreed above 200 Hz upto 200 kHz. The resonant peak at 300 kHz in the in-loop measurement did not appear in the out-of-loop measurement since the inherent $1/e^2$ linewidth of each tooth of the Ti:Sapphire OFC is narrower than 30 kHz (see Appendices 4.4).

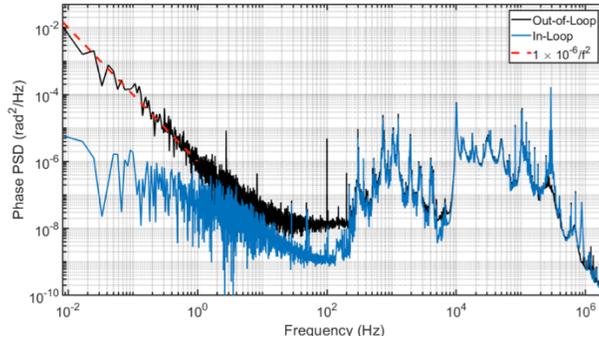

Fig. 5. PSD of the phase noise of the in-loop(blue) and out-of-loop(black) heterodyne beats between the copy of the clock laser and the OFC. The phase PSD of the out-of-loop measurement showed $1\times10^{-6}/f^2$ dependence below 10 Hz.

The accumulated phase in the out-of-loop heterodyne beat integrated from 8.3 mHz to 1.95 MHz was 0.5384 rad, mostly originating the components with frequencies higher than 200 Hz. This small phase accumulation < 1 rad indicates that the residual linewidth of the optical frequency comb with respect to the copy of the clock laser was smaller than 8.3 mHz. The linewidth can be estimated by extrapolating the low frequency components to $1\times10^{-6}/f^2$. The low cutoff frequency when the total accumulated phase becomes 1 rad can be referred as the residual linewidth. The estimated residual linewidth was 1.4 µHz. This is negligibly small compared to the sub-10 mHz linewidth of the current state-of-the-art CW laser [11]. Therefore, the linewidth of the stabilized frequency comb, which is the square-sum of the linewidth of the optical standard and of all of the PLLs from the remote site, would be comparable to the linewidth of the optical standard itself, demonstrating the direct phase locking of our Ti:Sapphire OFC to the remote optical frequency standard, avoiding the remnant fiber-induced phase noises.

## 2.3 Total relative instability of the OFC

With both of the $f_{CEO}$ and the $f_{rep}$ stabilized tightly, the OFC produced optical signals from 500 nm to 1,100 nm, whose frequencies were as precisely determined as the optical standard clock laser itself. Our Ti:Sapphire OFC could be phase-locked to an optical reference without a phase slip for more than 40 h [41]. In order to investigate the long-term relative stability of the OFC with respect to the optical standard delivered through an optical fiber network, the frequencies of the heterodyne beat signals at each optical PLL were measured using Λ-type resolution-enhanced frequency counters in continuous mode (Agilent 53220A and 53230A) (Fig. 6). All instabilities show $\tau^{-1/2}$ dependence that is characteristic to white frequency modulation. The values at 1 s and 1,000 s for each PLL are summarized in Table 2.

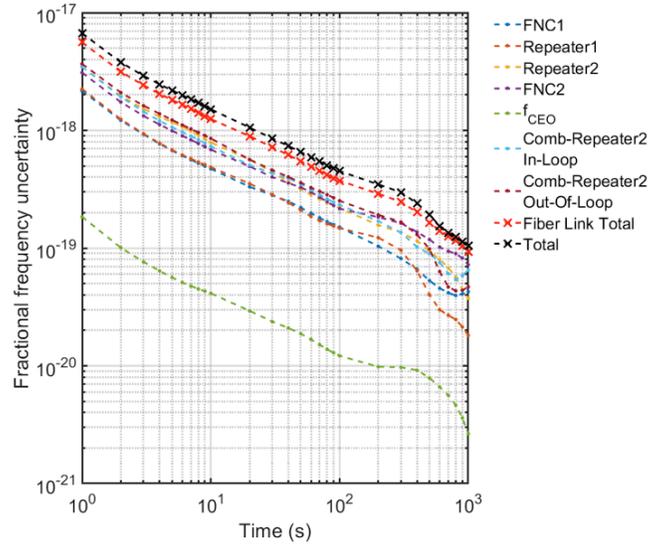

Fig. 6. Fractional frequency instabilities of the phase-locked heterodyne beat frequencies measured by Λ-type frequency counters with the measurement interval of 1 s. All measurements showed a $\tau^{-1/2}$ dependence, indicating that white frequency modulation was the major measurement noise. The total uncertainty of the remote OFC to the optical standard is $6.7(1)\times10^{-18}$ at 1 s and $1.05(3)\times10^{-19}$ in 1,000 s including the fiber link.

Table 2: Fractional frequency instability of the beat signal for each PLL.[a]

| | σ (1 s) | σ (1000 s) |
|---|---|---|
| FNC1 | $2.17(3) \times 10^{-18}$ | $4.3(1) \times 10^{-20}$ |
| Repeater1 | $2.25(3) \times 10^{-18}$ | $1.81(3) \times 10^{-20}$ |
| Repeater2 | $3.46(5) \times 10^{-18}$ | $3.7(1) \times 10^{-20}$ |
| FNC2 | $3.09(5) \times 10^{-18}$ | $7.2(2) \times 10^{-20}$ |
| $f_{CEO}$ | $1.84(3) \times 10^{-19}$ | $2.65(8) \times 10^{-21}$ |
| Repeater2 – OFC (In-Loop) | $3.47(5) \times 10^{-18}$ | $6.5(2) \times 10^{-20}$ |
| Repeater2 – OFC (Out-of-Loop) | $3.67(6) \times 10^{-18}$ | $4.7(1) \times 10^{-20}$ |
| Fibre-link Total | $5.60(8) \times 10^{-18}$ | $9.4(3) \times 10^{-20}$ |
| Fibre-link & OFC Out-of-Loop | $6.7(1) \times 10^{-18}$ | $1.05(3) \times 10^{-19}$ |

[a] Measured with Λ-type frequency counters. The measurement interval was 1 s. Fractional frequency instabilities at averaging times of 1 s and 1000 s are indicated. The 1σ statistical errors are indicated in parentheses.

The difference between the two values for FNCs originated from the differences of their beat frequencies (80 MHz/20 MHz), carrier frequencies (optical frequencies, 698 nm/1,397 nm), type of the optical fiber (polarization-maintaining or otherwise), optical fiber length (100 m/2.9 km), and the use of a repeater laser. The different carrier frequencies for the repeater phase locks were the major contributor to the instability discrepancies between Repeater1 and Repeater2. The OFC was phase-locked to the copy of the clock laser with the instability of $3.67(6) \times 10^{-18}$ at 1 s and $4.7(1) \times 10^{-20}$ at 1,000 s. Since all the instabilities were independent, the total instability can be calculated by the root sum square of all the components. The obtained total relative fractional frequency instability of the optical frequency comb with respect to the remote optical standard was determined to be $6.7(1) \times 10^{-18}$ at 1 s and $1.05(3) \times 10^{-19}$ at 1,000 s including the fiber-link. This result transcends the instability of the current state-of-the-art optical atomic clocks, assuring that all frequency modes of the OFC can have the same order of instability to the optical atomic clocks with the scheme presented here. Since 70% of the Ti:Sapphire oscillator output power remains unused in the present study, an extension to the shorter wavelength region can be readily achievable. Distribution of the optical frequency standard and direct coherence transfer demonstrated here thus pave the way for conducting ultrahigh-precision spectroscopy in the visible and ultraviolet regions at distant places where telecommunication optical fiber networks are available.

## 3. Conclusion

The direct phase-locking of a Ti:Sapphire OFC to a Sr optical lattice clock laser was demonstrated through a 2.9-km-long ubiquitous optical fiber network consisting of multiple switches and two near-infrared repeater lasers. The relative fractional frequency instability of the stabilized OFC with respect to the optical standard clock laser was $6.7(1) \times 10^{-18}$ at 1 s and $1.05(3) \times 10^{-19}$ at 1,000 s including the fiber-link, which is compatible with today's best optical frequency standards. This result indicates that a Ti:Sapphire OFC can directly duplicate the coherence and the instability of the optical standard via optical fiber networks of intra-city or on-campus scales. This setup can be easily scalable for connecting multiple laboratories through fiber networks, and it opens the door for pioneering spectroscopy experiments at various wavelengths with the precision of optical atomic clocks, unrestrained by geographical access to next-generation time and frequency optical standards. Tests of physics beyond the current paradigm represented by the Standard Model and general relativity could now be propelled by the global comparison of optical frequencies with this unprecedented precision.

**Funding**
Advanced Photon Science Alliance (APSA) in the Photon Frontier Network Program of MEXT, Japan; JSPS KAKENHI Grant (JP17H06205, JP17944928).

**Acknowledgments**
The authors thank H. Katori for providing access to the clock laser and other equipment in his laboratory. The authors appreciate the technical support provided by T. Takano, N. Ohmae, T. Akatsuka and I. Ushijima. The authors thank M. Kuwata-Gonokami for presenting the conceptual idea in Advanced Photon Science Alliance (APSA) project that led to the present work, and providing basic equipment at the OFC site.

## 4. Appendices

*4.1 Phase-locked loop (PLL)*

The heterodyne beat was detected by a PD and amplified by 40 dB using a home-made RF amplifier. In order to compensate for the small power variation in the beat signal, an auto-gain control amplifier was employed after the 40 dB amplifier. After a band-pass filter (Mini-Circuits, BBP-10.7+ or BBP-21.4+, depending on the locking frequency), the signal was sent to a phase-frequency detector that generated the error signals for the feedback by comparing the frequency of the signal and the reference frequency from a function generator (Tektronix, AFG1022). The function generator was externally referenced to a GPS-disciplined 10 MHz Rb clock. The error signal was fed to a home-built PID servo controller where the feedback signal was generated. The feedback signal was sent to appropriate adjustable components of each stabilization loop: voltage-controlled crystal oscillators for the FNC, PZTs and driving currents of for the ECDLs, the pump power of the Ti:Sapphire oscillator for controlling $f_{CEO}$ and the PZTs attached to the laser cavity for the phase stabilization of the frequency comb to the clock laser.

*4.2 Estimation of the residual one-way fiber phase noise in PLLs*

The residual one-way fiber phase noise after the round-trip phase noise was cancelled was estimated by expanding the discussion of Ref. 30. We introduced the remaining in-loop phase noise of the PLL in contrast to Ref. 30 where a perfect PLL was assumed.

$$\delta\phi_{IL}(t) = \delta\phi_{Fiber,RT}(t) - \phi_c(t) + \phi_c(t - 2\tau)$$

$\delta\phi_{IL}(t)$, $\delta\phi_{Fiber,RT}(t)$, $\delta\phi_c(t)$ are the remaining in-loop phase noise, the round-trip free-running fiber phase noise, and the phase correction for cancelling the fiber noise. $\tau$ is the travel time of the light through a fiber-link with length $L$. $\delta\phi_{Fiber,RT}(t)$ and $\delta\phi_{Fiber}(t)$, the one-way free-running fiber phase noise, can be expressed as

$$\delta\phi_{Fiber,RT}(t) = \int dz \left[\delta\phi_z\left(t - \frac{z}{v}\right) + \delta\phi_z\left(t - 2\tau + \frac{z}{v}\right)\right] \sim \int dz[2\delta\phi_z(t) - 2\tau\delta\phi'_z(t)]$$

$$\delta\phi_{Fiber}(t) = \int dz \left[\delta\phi_z\left(t - \tau + \frac{z}{v}\right)\right] \sim \int dz \left[\delta\phi_z(t) - \tau\delta\phi'_z(t) + \frac{z}{v}\delta\phi_z'(t)\right]$$

where $\delta\phi'_z(t) = \frac{d\phi_z(t)}{dt}$ and $v$ is the light phase velocity in the fibre [30]. Using the above equations, the residual one-way fiber phase noise after the phase correction becomes as below.

$$\delta\phi_{one-way,res}(t) = \int dz \left[\frac{z}{v}\delta\phi'_z(t)\right] - \frac{1}{2}\delta\phi_{IL}$$

The PSD of the residual one-way fiber phase noise therefore consist of the terms from the free-running phase noise, the remaining in-loop noise, and the crossterm of the two.

When a repeater laser is in use in FNC, the residual phase noise of the repeater laser should be considered as well. The round-trip free-running fiber noise including the repeater becomes as below.

$$\delta\phi_{Fiber,RT}(t) = \int dz \left[\delta\phi_z\left(t - \frac{z}{v}\right) + \delta\phi_z\left(t - 2\tau + \frac{z}{v}\right)\right] + \delta\phi_{repeater}(t - \tau)$$

Following the same analysis procedure described above, one can attain the PSD of the residual one-way fiber noise that consists of the terms from the free-running fiber noise, the remaining in-loop phase noise, the phase noise of the repeater laser, and the crossterms of the three. The round-trip free-running phase noise of FNC1 and FNC2 are shown in Fig. 7.

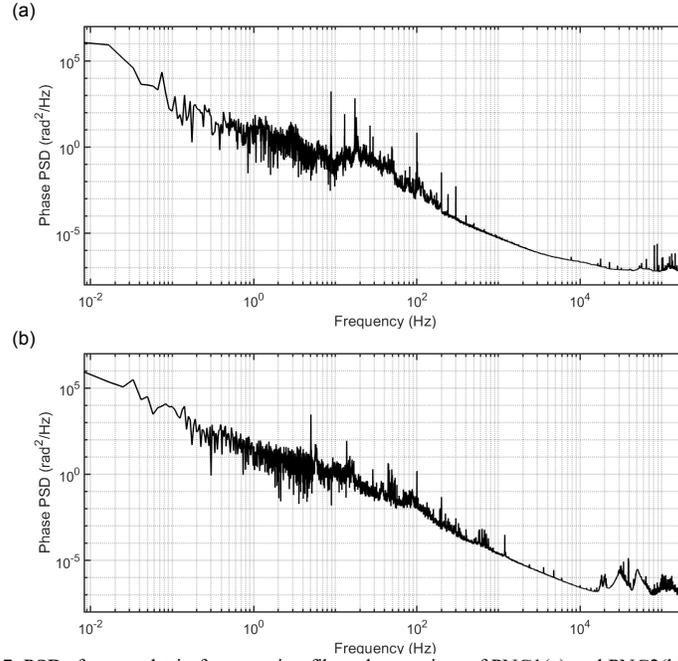

Fig. 7. PSDs for round-trip free-running fiber phase noises of PNC1(a) and PNC2(b). They fall off as $1/f^2$ indicating white frequency modulation noise.

## 4.3 Stabilization of $f_{CEO}$

$f_{CEO}$ of the OFC was stabilized using the self-referencing technique with an $f$-$2f$ interferometer. The inherent linewidth of the home-built Ti:Sapphire OFC's $f_{CEO}$ was in the order of kilohertz, as shown in Fig. 8(a). The narrow free-running linewidth was one of the key aspects for achieving the unprecedented optical frequency comb instability in this work.

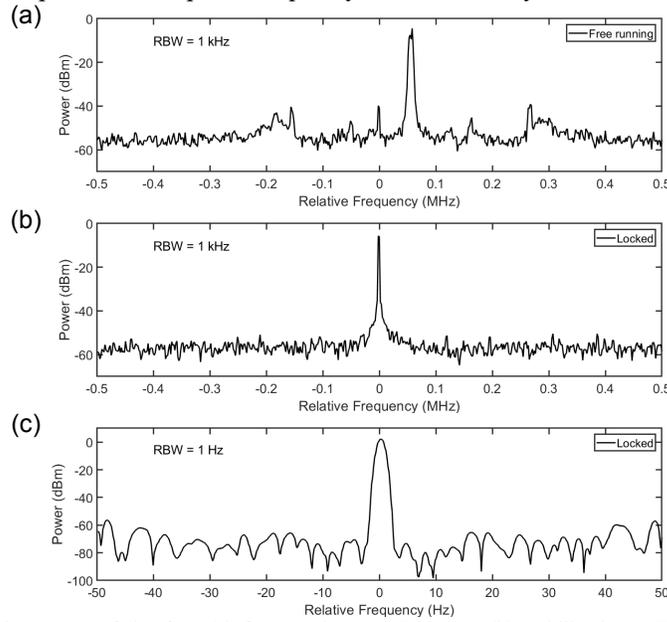

Fig. 8. Spectra of the $f_{CEO}$: (a) free-running, 1 MHz span, (b) stabilized, 1 MHz span, (c) stabilized, 100 Hz span. Spectral components at around 200 kHz from the center frequency of the $f_{CEO}$ were from noises of the pump laser.

## 4.3 Stabilization of $f_{CEO}$

As shown in Fig. 9, the $1/e^2$ linewidth of the free-running OFC used in this experiment spanned about 20 kHz. The narrow linewidth of the OFC ensured that the broad pedestal phase noise components above 200 kHz of the repeater lasers (Fig. 2(a) and 2(c)), did not affect to the stability of the optical frequency comb.

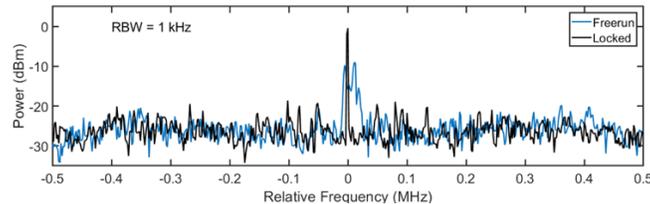

Fig. 9. Spectrum of the beat signal between the clock laser copy and the optical frequency comb. The blue line shows the spectrum when the comb is not stabilized to the clock laser. The central feature over the width of about 20 kHz indicated the convoluted linewidth of the clock laser and the optical frequency comb. Due to the narrow linewidth of the clock laser, the linewidth of the optical frequency comb dominated the observed linewidth. The spectrum of the beat signal when the frequency comb was phase-locked to the clock laser is shown in black.